\documentclass[useAMS,usenatbib]{mn2e}
\usepackage{epsfig,amsmath,natbib}
\usepackage{color}

\def\be{\begin{equation}} 
\def\ee{\end{equation}}

\def\kms{\,{\rm {km\, s^{-1}}}} 
\def\cc{\,{\rm {cm^{-3}}}} 
\def\msun{{\Msun}}

\def\HH{${\rm {H_2}}\,\,$}

\def\gsim{\lower.5ex\hbox{\gtsima}} 
\def\lsim{\lower.5ex\hbox{\ltsima}} \def\gtsima{$\; \buildrel > \over 
\sim \;$} \def\ltsima{$\; \buildrel < \over \sim \;$} \def\prosima{$\; 
\buildrel \propto \over \sim \;$} \def\gsim{\lower.5ex\hbox{\gtsima}} 
\def\lsim{\lower.5ex\hbox{\ltsima}} 
\def\simgt{\lower.5ex\hbox{\gtsima}} 
\def\simlt{\lower.5ex\hbox{\ltsima}} 
\def\simpr{\lower.5ex\hbox{\prosima}} \def\la{\lsim} \def\ga{\gsim} 
  
 \def\gtsima{$\; \buildrel > \over \sim \;$} 
\def\ltsima{$\; \buildrel < \over \sim \;$} 
\def\gsim{\lower.5ex\hbox{\gtsima}} 
\def\lsim{\lower.5ex\hbox{\ltsima}} 
\def\simgt{\lower.5ex\hbox{\gtsima}} 
\def\simlt{\lower.5ex\hbox{\ltsima}} 
\def\simpr{\lower.5ex\hbox{\prosima}} 
\def\la{\lsim} 
\def\ga{\gsim}

\def\msun{\,{\rm \Msun}}

\def\E3{{\cal E}_{\rm g}^{III}}

\def\Msun{\rm M_\odot}

\def\Tvir{T_{vir}} 
\def\r12{r_{1/2}} 
\def\x12{x_{1/2}} 
\def\v12{v_{1/2}}

\title[SMBH in galaxy mergers]{Can supermassive black hole seeds form in galaxy mergers?} 
\author[Ferrara et al.]{A. Ferrara$^{1,5}$, F. Haardt$^{2,3}$, R. Salvaterra$^4$ \\
$^{1}$ Scuola Normale Superiore, Piazza dei Cavalieri 7, I-56126 Pisa, Italy\\
$^2$ DiSAT, Universit\`a degli Studi dell'Insubria, via Valleggio 11, I-22100 Como, Italy\\
$^3$ INFN, Sezione di Mlano-Bicocca, Piazza della Scienza 3, I-20126 Milano, Italy\\
$^4$ INAF/IASF-MI, via Bassini 15, I-20133, Milano, Italy\\
$^5$ Kavli Institute for the Physics and Mathematics of the Universe (WPI), Todai Institutes for Advanced Study, the University of Tokyo\\
}

\begin{document} 
 
\date{\today} 
 
\pagerange{\pageref{firstpage}--\pageref{lastpage}} \pubyear{2012} 
 
\maketitle 
 
\label{firstpage} 
\begin{abstract} 
It has been recently suggested that supermassive black holes at $z\approx 5-6$ might form from super-fast ($\dot M \simgt 10^4 M_\odot$ yr$^{-1}$) accretion occurring in unstable, massive nuclear gas disks produced by mergers of Milky--Way size galaxies.   Interestingly, such mechanism is claimed to work also for gas enriched to solar metallicity. These results are based on an idealized polytropic equation of state assumption, essentially preventing the gas from cooling. We show that under more realistic conditions, the disk rapidly $(< 1 $ yr) cools, the accretion rate drops, and the central core can grow only to $\approx 100 M_\odot$. In addition, most of the disk becomes gravitationally unstable in $\approx 100$ yr, further quenching the accretion. We conclude that this scenario encounters a number of difficulties that possibly make it untenable.
\end{abstract}


\section{Motivation}
\label{Mot}
The origin of the supermassive black holes (SMBH) we now routinely observe at epochs within the first cosmic Gyr represents one of the most intriguing puzzles in structure formation.  The current paradigm implies that these objects have gathered their mass by accreting the surrounding gas onto a much smaller seed BH  (\citealt{Volonteri03}; \citealt{Volonteri05}; \citealt{Lodato06}; \citealt{Volonteri03}; \citealt{Natarajan11}; \citealt{Tanaka09}; \citealt{diMatteo08}; \citealt{Li07}). As massive stars end their evolution into BHs of mass $m_0 \approx 10-50 \msun$ this mechanism provides the most natural route to produce the initial seeds. However, this scenario has to face at least two, partly related, serious difficulties. First, in order to reach the typical SMBH mass ($\approx 10^{8-9} \msun$) in the limited time ($\approx$ Gyr) available up to $z=6$ the accretion must not only always proceed at the Eddington rate, but also possibly with an unusually low radiation efficiency. However, several studies (e.g. \citealt{Alvarez08}, \citealt{Milosavljevic09}) have now shown that stellar BHs are actually characterized by very low ($\dot M\approx 10 ^{-12} M_\odot$ yr$^{-1}$) accretion rates due to radiative feedback and because they spend most of their lifetime in low-density regions. 

These difficulties would be greatly smoothed out by a larger $m_0 \simgt 10^3 \msun$ seed mass. It is then worth exploring viable formation paths for these intermediate mass SMBH seeds. Long before these problems were realized, proposals for the production of more massive ($m_0 \approx 10^{4-6} \Msun$) seeds were made (\citealt{Loeb94}; \citealt{Eisenstein95}) which have now developed into more complete scenarios (\citealt{Begelman06}; \citealt{Shang10}; \citealt{Johnson12}; \citealt{Regan09}; \citealt{Petri12}). This channel invokes the formation of massive black hole seeds in environments where gas gravitational collapse proceeds at very sustained rates, $\dot M_g \simgt 0.1 -1\Msun$ yr$^{-1}$, i.e. about 100 times larger than for standard metal-free star formation; these objects are often dubbed as ``direct collapse black holes'' (DCBH) to distinguish them from the smaller seeds of stellar origin discussed above. Where are these environments to be found? So far, the most promising candidates are dark matter halos with virial temperature $\Tvir \simgt 10^4$ K. In these halos the primordial gas radiatively cools via collisional excitation of the hydrogen $1s \rightarrow 2p$ transition followed by a Ly$\alpha$ photon emission. Given the strong temperature sensitivity of such process, the gas collapses almost isothermally, $1+d\ln T/d\ln \rho \equiv \gamma \approx 1$, thermostating the temperature at $T\approx 8000$ K. Under these conditions, gas fragmentation into sub-clumps is almost completely inhibited (\citealt{Schneider02}; \citealt{Omukai05}; \citealt{Omukai08}; \citealt{Cazaux09}) and collapse proceeds to very high densities unimpeded.

Even this scenario is not free from concerns. In fact, it requires that a sufficiently strong Lyman-Werner UV radiation field is present to prevent \HH molecule formation and the subsequent rapid cooling. Similar enhanced cooling, leading to fragmentation of the gas, can also be produced by a non-negligible heavy element abundance; however, is it not clear if relatively large (2-3 $\sigma$ density fluctuations) \textit{unpolluted} halos can form. 

As an alternative route to form a massive seed, \citet{Mayer10} and \citet{Bonoli12} (but see also similar ideas put forward by  \citet{Begelman08}, \citet{Begelman10} and \citet{Ball12} 
discussing the evolution of ``quasi-stars'') noted that merger-driven 
gas inflows produce an unstable, massive nuclear gas disk. Accretion from this disk feeds a central core, which, according
to these studies might grow up to $10^8 \msun$ in a very short time ($\simlt 10^5$ yr). As this central core becomes Jeans unstable,
it might lead to the direct formation of a SMBH \textit{even for a solar metallicity gas}. If the extremely high accretion rates $\dot M \simgt 10^4 \msun$ yr$^{-1}$ required can be sustained is a question that needs more scrutiny. Although attractive, we show in the following that this scenario might encounter a number of difficulties that possibly make it untenable.


\section{Nuclear  disk properties}
\label{Nuc}
The SPH simulation of \cite{Mayer10} follows the evolution of the merger of two high redshift, still well-formed, disk galaxies embedded 
in a dark matter halo of mass $M=10^{12} \Msun$. As a result of the merger a nuclear, self-gravitating disk 
of radius $r_d\approx 40$ pc and mass $M_d=2\times 10^{9} \Msun$ forms. The disk gas is highly turbulent, with a velocity dispersion $\sigma \approx 100 \kms$; the turbulent energy is ultimately drained from the gravitational energy of the system
driving the collision first, and inducing non-axisymmetric instabilities and spiral arms later on. The disk orbital period 
at 20 pc is $5 \times 10^4$ yr. 

The simulation shows that the disk gas efficiently loses angular momentum and is transported towards the center with 
astonishingly high rates, $\dot M > 10^4 \Msun$ yr$^{-1}$, where it accumulates in a pc-sized, roughly spherical structure 
(the core), which therefore grows to 13\% of the total disk mass, $2.6\times 10^8 \Msun$, in $\simeq 0.1$ Myr. Although the simulation was stopped at that time, \cite{Mayer10}
suggested that this core structure is likely to evolve into a central black hole surrounded by an accreting envelope, i.e., a 
quasi-star as described by \cite{Begelman08}. This guess is essentially based on the fact that the central core temperature, $T_c \approx
10^7$ K, is so high to prevent any fragmentation and subsequent star formation in the gas on its way to the newly formed compact object.

It is clear that the very high temperature of the gas is instrumental in keeping the accretion rate as large as observed in the  simulation. This can be easily deduced from simple Jeans argument, the accretion rate being  
$\dot M \approx M_J/t_{\rm ff}$, where $M_J$ is the Jeans mass and $t_{\rm ff}$ is the free-fall timescale. Numerically 
\be
\dot M \approx \frac{\pi^2}{8G} c_s^3 \simeq 5\times 10^3  \left(\frac {T}{10^7 \textrm{K}}\right)^{3/2} M_\odot {\rm yr}^{-1},
\label{mdot} 
\ee 
where here $c_s$ is the disk gas sound speed. It is then straightforward to conclude that the accretion rate is so high because of  the high ``effective'' temperature of the gas ($\approx  10^7$ K).

A similar result is obtained considering a disk-like accreting flow. In this case 
$\dot M = 3\pi \nu \Sigma$, where $\nu= \alpha c_s H = \alpha c_s^3 /\pi G \Sigma$ 
is the turbulent viscosity for a thin disk, $H$ the disk 
scale height, and $\Sigma$ the disk surface density. We find $\dot M = 3\alpha (c_s^3/G) \approx 10^3 
(T/10^7 {\rm K})^{3/2}$ $M_\odot$yr$^{-1}$ for the usually assumed value $\alpha = 0.1$. 

In order to study the properties of the accreting flow, we need to estimate its density. 
The density profile for an assumed isothermal disk \citep{Spitzer42} is
\be
\rho(r,z) = \frac{\Sigma(r)}{2H(r)} \textrm{sech}^2\left(\frac{z}{H(r)}\right)
\label{profile} 
\ee 
where the scale height $H$ is given by:
\be
H = \frac{c_s^2}{\pi G \Sigma} 
\label{height} 
\ee 
We can then define a characteristic density at any given radius by weighting the density profile over the column density: 
\be
\langle\rho\rangle  = \mu m_p \langle n \rangle = \frac{1}{\Sigma} \int_{-\infty}^\infty{\rho^2 dz}, 
\label{density} 
\ee 
Here $m_p$ is the proton mass and $\mu=0.65$ is the mean molecular weight of a gas 
with solar abundances. In the \cite{Mayer10} simulation the mass surface density outside the 
central pc is found to be in the range $\Sigma(r> 1 \mathrm{pc)} =10^{4-8} \Msun$ pc$^{-2}$.
A midrange value, $\Sigma = 10^{6} \Msun$ pc$^{-2}$, gives $\langle n \rangle = 1.4 \times 10^6 \cc$. 

\section{Hot disks}

The direct collapse black hole scenario described by \cite{Mayer10} relies on a very strong assumption regarding the temperature evolution of the gas in the disk and central core. In fact, the authors adopted an equation of state (EOS) based on the work of \citep{Klessen07}, who  studied the interstellar medium in starburst galaxies. According to such results, an EOS 
$T\propto \rho^{\gamma-1}$, with $\gamma = 1.1-1.4$ was assumed. The precise value of the adiabatic index depends on the gas 
density, and it was set softer by \cite{Mayer10} in the densest regions (as, e.g., in the central core). This EOS embeds both the heat input from supernovae, and the energy losses via radiative processes. Both processes were not explicitly modeled in the nuclear disk/core.

\subsection{Analytical estimates}
 
Are such thermodynamical properties of the disk gas consistent with a more realistic energy equation 
and, in particular, with gas cooling? 

Let us start by evaluating the cooling time. This can be expressed as 
\be
t_c = \frac{3}{2} \frac{k_B T}{\langle n \rangle \Lambda(T,Z)}, 
\label{tcool} 
\ee 
where $k_B$ is the Boltzmann constant, and $\Lambda$ is the standard cooling function depending on
temperature and metallicity of the gas. We assume that at $T\approx 10^7$ K the dominant cooling mechanism
is bremsstrahlung (free-free)\footnote{At 
$Z=Z_\odot$ cooling due to metal line emission is actually 2.5 times higher than free-free; our argument 
is then a conservative, valid independently of gas metallicity.} , 
\be
\Lambda(T,Z) \approx \Lambda_{ff}(T) = \Lambda_0 T^{1/2}
\langle g(\nu,T) \rangle, 
\label{Lambda} 
\ee 
where $\langle g(\nu,T) \rangle = 1.24$ is the mean Gaunt factor appropriate for the temperatures of
interest here and $\Lambda_0 \simeq 1.43 \times 10^{-27}\textrm{erg cm}^{3} \rm s^{-1} \rm K^{-1/2}$. By assuming a fully ionized gas we obtain
\be
t_c \simeq 12.8 \left(\frac{T}{10^7 \textrm{K}}\right)^{1/2} 
\left(\frac{\langle n \rangle }{1.4\times 10^6 \cc}\right)^{-1} \textrm{yr}.
\label{tcoo1} 
\ee 
The previous calculation assumes that the gas is optically thin, which may not be the case. 
Indeed, the disk optical depth along the vertical
direction is $N_H = \Sigma / \mu m_p \sim 10^{26}$ cm$^{-2}$, corresponding to an
electron scattering optical depth $\tau = N_H \sigma_T \sim 100$,
where $\sigma_T \simeq 6.65\times 10^{-25} $ cm$^{2}$ is the Thomson cross section. Photons do not stream directly out of 
the disk, but perform a random walk and leak out of the disk on a diffusion time scale, $t_d = H \tau/c \sim
3\times 10^3$ yr. However, this time is a good order of magnitude shorter than the free-fall time, $t_{\rm ff} = \sqrt{3\pi/32 G \langle \rho \rangle}
\sim 4\times 10^4 {\rm yr} \gsim 10$ $t_d$, i.e. the gas will effectively cool, and fragment in low mass lumps that will start to orbit around the center. As a result, accretion onto the central core will be almost completely quenched. 

The simple argument above shows that the disk would dissipate its thermal energy in a diffusion time scale, unless some energy is injected in the gas, balancing radiative losses.  We can plausibly envisage two types of energy sources: (a) supernova
explosions occurring in the disk itself, or (b) gravitational energy. The first energy input is obviously 
associated with star formation activity; the latter originates from the dissipation of the kinetic energy 
of the collapsing gas. To prevent the fast cooling of the gas, 
both sources must inject energy at a rate equal to the dissipation rate,
\be
\dot E_d \equiv \vert\frac{dE}{dt}\vert \approx \left(\frac{M_d}{\mu m_p}\right) \frac{kT}{\max\{t_c,t_d\}} = \left(\frac{M_d}{\mu m_p}\right) 
\frac{kT}{H\tau} c,
\label{Ed} 
\ee 
that can be written as 
\begin{eqnarray}
\dot E_d &=&\left(\frac{\pi G \mu m_p c}{\sigma_T}\right) M_d \\ 
&=& 9\times 10^{46} \left(\frac{M_d}{2\times 10^9 \Msun }\right)\textrm{erg s}^{-1},
\label{Ed1} 
\end{eqnarray}
which is, as expected, comparable to the Eddington luminosity for a disk--like configuration..

As far as supernovae are concerned, we can estimate the energy input rate, $E_i^{\rm sn}$ as follows. Suppose that 
$\nu =0.01 \Msun^{-1}$ supernovae\footnote{Appropriate for a standard Salpeter IMF extending in the mass range 0.1-100 $\Msun$.} are produced for each solar mass 
of stars formed, injecting a fraction $\eta \approx 0.1$ of their total energy, $E_0=10^{51}$ erg, in thermal form. If we further 
define the star formation rate as $\psi$, the energy injection rate is
\be
\dot E_i^{\rm sn} =\eta \nu E_0 \psi = 3.2 \times 10^{40} \left(\frac{\psi}{\Msun {\rm yr}^{-1} }\right)\textrm{erg s}^{-1} , 
\label{Eisn} 
\ee 
implying that an unreasonably high star formation rate $\psi \simgt 10^6 \Msun {\rm yr}^{-1}$ would be required. It is also well 
possible that supernova feedback destroys the disk completely.
In any case, such a high star formation rate could be sustained only for a time $M_d/\psi = 2000$ yr (comparable to $t_d$) 
before the gas is completely consumed.
 
Alternatively one might argue that energy can be drained from the gravitational potential rather than being continuously supplied by star formation. If the gas is shock-heated during disk formation to the temperature required to guarantee the desired accretion rate, i.e. $T\approx 10^7$ K, we run into the strong requirements set by radiative energy dissipation. In fact, the disk gravitational energy is
\be
W  =\frac{2\pi}{3-2a} G \Sigma_0^2 r_0^3 \left[ \left(\frac{r_d}{r_0}\right)^{3-2a} -1\right] \simeq 4.5 \times 10^{57} \textrm{erg},
\label{Egrav} 
\ee 
where he surface density power-law $a=2.1$ has been determined by requiring that the integral of the surface density 
$\Sigma(r)=\Sigma_0 (r/r_0)^{-a}$, with $\Sigma_0=10^8 \Msun$ pc$^{-2}$, from the inner radius $r_0=1$ pc out to $r_d$ gives the correct disk mass $M_d$. As for the case of supernovae, we see that this gravitational energy would be radiated away on a very short time scale, $W/\dot E_d \approx 3500$ yr. 

\begin{figure}
\includegraphics[width=90mm]{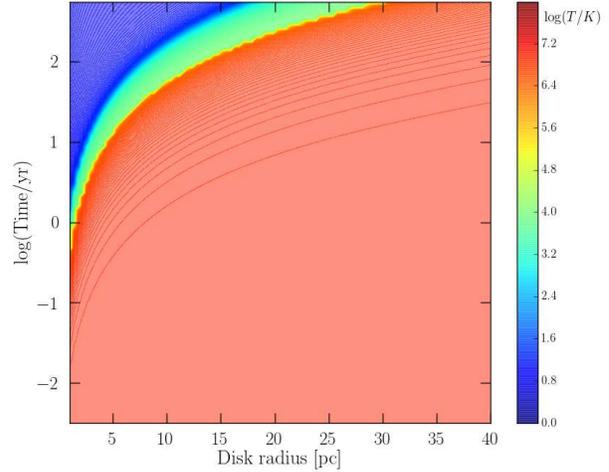}
\caption{Time evolution of the nuclear disk temperature as a function of disk radius, as indicated by the colorbar.} 
\label{Fig01}
\end{figure}
\begin{figure}
\includegraphics[width=90mm]{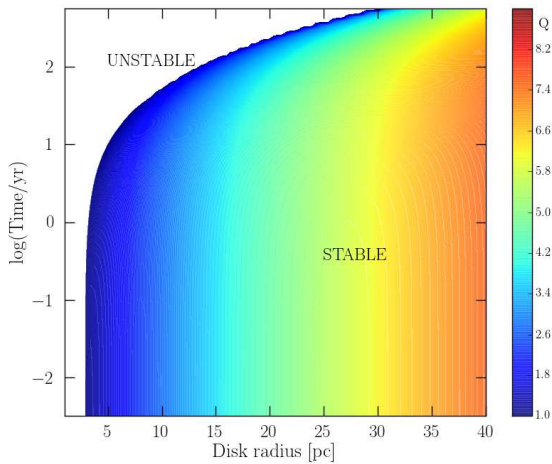}
\caption{Time evolution of the nuclear disk Toomre $Q$ parameter, as shown by the colorbar. White areas correspond to gravitationally unstable ($Q<1$) regions.} 
\label{Fig02}
\end{figure}

\subsection{Numerical solutions}

We are now interested in determining the detailed thermal evolution of the disk and the corresponding accretion 
rate evolution onto the central core. To this aim we write the energy equation for the disk gas:
\be
\frac{\partial}{\partial t} \left( \frac{3}{2} nkT\right) = - n^2 \Lambda(T,Z) p(\tau)+ \frac{1}{2}\nu \Sigma\frac{r^2}{H} \left(\frac{\partial \Omega}{\partial r} \right)^2, 
\label{energy} 
\ee
The rotation angular frequency is $\Omega(r)=v_\phi/r =\kappa_e /\sqrt{2}$, 
where $v_\phi(r)$ is the disk rotation velocity, and $\kappa_e(r)$ is the epicyclic frequency. The quantities $T, n, \Sigma$ depend on radius, which ranges from the value at the edge of the central core ($r_0=1$ pc) to the outer disk radius, $r_d=40$ pc. Thus the radial temperature profile at any given time is determined by radiative losses (first term on the r.h.s.) and viscous heating (second term).
However, we find that viscous heating is generally negligible with respect to energy cooling losses. We complement the above equation with the initial condition $T(r,t=0) \equiv \mathrm{const.} =10^7$ K, i.e., we assume that the disk has been initially heated by compressions and shocks following the merger between the two galaxies. The function $p(\tau$) 
takes into account the fact that the disk is not optically thin to cooling radiation, and can be identified with the average 
escape probability of photons from a slab of vertical optical depth $\tau=N_H\sigma_T$. 
The following approximation is sufficient for our purposes: 
\be
p(\tau) \simeq \frac{1-{\rm e}^{-\tau}}{\tau}.
\label{escape} 
\ee

In our study we considered isochoric solutions to eq. \ref{energy}, in which therefore the density is independent of time
(but has a dependence on radius set by eq. \ref{density}). This is justified by the fact that in the disk $t_c \ll H/c_s$: under these conditions pressure cannot be restored sufficiently rapidly by shock waves and the gas cools at almost constant density. 

The gas cools more rapidly in the inner disk regions where the density is higher; at the same time it accretes  onto the central core
at a rate set by the temperature at the $r_0$ boundary ($\dot M \propto c_s^3$). 
The build-up of the central core to masses larger than the Jeans mass, $M_J$, leading to the final collapse to a black hole seed is however hampered by two facts. First, the core growth becomes slower as the disk gas progressively cools. Second, and at the same time, the inner regions of the disk become gravitational unstable and fragment. Let us analyze these two occurrences in more detail.   

The time evolution of the nuclear disk temperature as a function of the disk radius, $r$, obtained from the numerical solution of eq. \ref{energy} is shown in Fig. \ref{Fig01}. From there we see that in $< 10$ yr, the central regions of the disk, within $r=5$ pc, have already cooled down to $\simlt 10^4$ K. Due to the decreasing density of the more external regions, these remain hot for a longer time; however after $\approx 10^3$ yr, the entire disk has cooled down. 

The cooling might also trigger (or amplify) the disk gravitational instability, usually identified by values of the Toomre parameter $Q<1$. This parameter can be written as  
\be
Q (r) =\frac{\sigma \kappa_e(r)}{\pi G \Sigma(r)}
\label{Q} 
\ee 

The disk rotation velocity at radius $r > r_0$ can be easily obtained by inserting the expression for $\Sigma$ used in eq. \ref{Egrav} in the following equation:
\begin{eqnarray}
&& v_\phi(r) =\sqrt{\frac{GM(r)}{r}}\\ 
&& = \left\{\frac{2\pi G\Sigma_0 r_0^2}{2-a} \left[\left(\frac{r}{r_0}\right)^{2-a} -1 \right] + G M_c\right\}^{1/2}  r^{-1/2}. \nonumber  
\label{vphi} 
\end{eqnarray}
Note that we have added the gravitational effects of the central core, assumed to have a mass $M_c= 0.13 M_d$ (see Sec. \ref{Nuc}), as found by \citet{Mayer10}. Although not self-consistent with our model, this assumption minimizes the fragmentation probability by providing an upper limit to the disk rotation velocity. The dependence of $Q$ on time and radius is reported in Fig. \ref{Fig02}. The very inner parts ($r \le 2-3$ pc) are born unstable\footnote{\cite{Hopkins13} notices 
that in turbulent disks  fragmentation can occur also for $Q>1$. This is due to the broad spectrum of stochastic density fluctuations that can produce rare but extremely high-density local mass concentrations that will easily collapse.}, i.e. they are prone to fragmentation already at $t=0$ as a result of their large surface density $\Sigma$. The disk fragmentation wave (i.e. the white region where $Q<1$ in the Figure) travels towards larger radii. Behind the wave the gas flow fragments in clumps orbiting the central core without falling onto it (at least not on the short time scale $\approx t_c(r_0)$ required). After about 100 yr the disk region within $0.5 r_d = 20$ pc has become unstable, while the outer has $Q \approx 6$. 

In spite of the low $Q$ values, the role of fragmentation in quenching the accretion flow onto the core is probably sub-dominant. The reason is that fragmentation occurs on the free--fall timescale. At $r=r_0$ we find that $t_{\rm ff} = 161$ yr. Such timescale is much longer than the cooling timescale, $t_c=0.37$ yr, i.e., the gas cools well before the disk fragments. Stated differently, it is the energy loss by radiation that quenches $\dot M$ rather than fragmentation which appears only at a later evolutionary stage, when the gas already cooled down. 

\begin{figure}
\includegraphics[width=86mm]{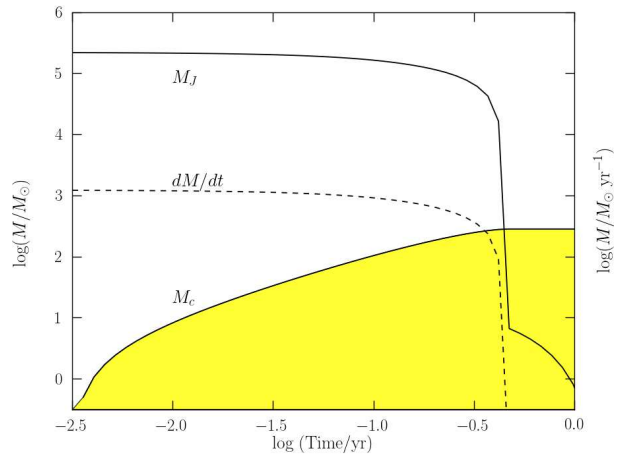}
\caption{Time evolution of the central core mass, $M_c$, Jeans mass, $M_J$, and accretion rate $dM/dt$ as indicated by the labels.} 
\label{Fig03}
\end{figure}

The challenge for the formation of the black hole seed as envisaged by \citet{Mayer10} lies in forming a sufficiently massive, Jeans unstable central core before accretion is quenched by gas cooling. Fig. \ref{Fig03} shows that this is
extremely difficult. The central core grows rapidly as it is fed by a very high initial accretion rate ($\dot M = 1-2\times 10^3 
M_\odot$ yr$^{-1}$) and reaches a mass of about $\simgt 100 M_\odot$ after about 4 months. Up to that point the core is still
gravitationally stable, as $M_J$ ia approximately 3 orders of magnitude larger. However, shortly after this phase, fast gas cooling induces a sudden drop both of $\dot M$ and $M_J$. The implication is that the core stops growing and starts to collapse. The evolution of a cold ($T\simlt 100$ K), collapsing, metal-enriched cloud has been subject to extensive studies in the recent years (\citealt{Schneider06}; \citealt{Bromm11} and references therein). All studies concur that the endpoint of the evolution is a large number of sub--solar mass clumps. As suggested by \citet{Omukai08},  such low--mass clumps might eventually result in a dense cluster  of low-- and intermediate--mass stars. 

The above arguments are similar (albeit relative to larger scales) to those given by \citet{Levin07} and \citet{Goodman03}, who suggested that disks in AGNs cool and fragment into stellar disks on scales much smaller than a parsec. The rapid cooling of the disk on scales of tens of pc could have consequences for gas accretion onto a SMBH {\it already} present in the galactic center. Indeed, several studies make the assumption that accretion at the resolution limit of the simulation translates into accretion onto the central SMBH. \citet{Dotti07} and \citet{Maio13} ran a set of simulations including gas cooling, and found that the accretion at the resolution limit (1 pc) was not significantly affected (the accretion rate would be, in any case, limited to $\lsim 0.01$ M$_\odot$/yr), though this result may depend upon the details of fragmentation vs. star formation.

In conclusion, in the merger+nuclear disk scenario as proposed by \cite{Mayer10}, the formation of black hole seeds as massive as $10^6-10^8$ $M_\odot$ appears problematic.

\section{Cold disks}

As a final possibility we explore the case in which the disk is formed in a cold, rather than hot, state. The collapse would initially induce bulk motions and turbulence (as indeed observed in the simulations). The initial velocity dispersion of the gas is $\sigma = v_c/\sqrt{3} =86 \kms$, where $v_c = 150 \kms$ is the virial velocity of the $10^{12} \Msun$ host halo at $z=7$. As the gas pressure is dominated by turbulence, we can estimate the infall rate from eq. \ref{mdot} by substituting $c_s$ with $\sigma$. This gives a much lower accretion rate, $\sim 200 \,\Msun {\rm yr}^{-1}$. 

However even this situation may not last for a long time, as turbulence can be dissipated quite efficiently, both in the supersonic and subsonic regimes \citep{MacLow99}.  Indeed, the ratio of the decay time of turbulence $t_{\rm td}$ to the free--fall time of the gas has been shown to be 
\be
\zeta(\kappa) =\frac{\kappa}{{\cal M}}\frac{1}{4\pi \eta_v} 
= \sqrt {\frac{32}{9\pi^2}} \frac{1}{4\pi \eta_v} = \textrm{const.} \simeq 0.7,
\label{turb} 
\ee 
where ${\cal M} = \sigma/c_s$ is the r.m.s. Mach number, $\kappa$ is
the ratio of the driving wavelength, of the order of the disk scale height $H$, to the Jeans
wavelength; $\eta_v=0.21/\pi$ is a (numerically calibrated) constant. To evaluate eq. \ref{turb} we have used $H (\sigma)$ from eq. \ref{height}. Note
that $\zeta$ is independent on the assumed gas temperature and $\sigma$, as long as the disk vertical support is provided by
turbulent pressure, as one might have suspected. 

Hence turbulence dissipation is not the major hampering factor for the central core growth; this is instead represented by
the fact that turbulence in a given fluid element of the disk is dissipated on a time scale much shorter that the time
necessary for the same element to reach $r_0$, i.e. the crossing--time of the disk from radius $r$: 
\be
t_\times(r) = \int_{r_0}^r \frac{dr'}{v_r(r')} = \frac{GM(r)}{3\alpha \sigma^3}.
\label{tc} 
\ee
In the previous expression $v_r$ is the radial velocity of the accreting material, $v_r=2\nu / 3 r$.  
We find that $t_{\rm td}$ ranges from about 70 yr at $r\approx r_0$ to
about $10^5$ yr at the disk outer edge; the ratio $t_{\rm
  td}/t_\times$ is found to be very small, i.e. $3 \times 10^{-5} <
t_{\rm td}/t_\times < 4 \times 10^{-3}$ in the same radial
range. Thus, turbulence is dissipated very quickly in comparison to
the accretion time scale; as a result, the accretion rate $\propto
\sigma^3$ also drops precipitously, making the black hole formation
scenario proposed unlikely. At the same time, the decreased level of
turbulent support leads to a disk gravitational instability and hence
to vigorous gas fragmentation.
The cold disk scenario bears some resemblance with the \citet{Begelman09} proposal that in turbulent
disks non-axisymmetric instabilities can funnel gas at the center through nested bars while fragmentation is suppressed by finite
disk thickness effects. It is unclear if this scenario applies to post-merger nuclear disks; however, more recently,
\citet{Hopkins13} has shown that turbulent disks are inevitably prone to fragmentation as a result of density inhomogeneities.

\section{Conclusions}
The large accretion rates required to form a massive black hole by preventing fragmentation of a metal-enriched gas
require that either the disk is heated at temperatures $\approx 10^7$ K, or that the disk can be initially set up 
in a cold and highly turbulent state. Both hypotheses are prone to serious problems. In the first 
case, the thermal energy is carried away very rapidly by cooling radiation, even considering the large optical depth
of the disk. In addition, the requirements in terms of either
supernova or gravitational energy to sustain such disk thermal budget
are truly enormous, and almost not plausible. We have shown that, under realistic thermodynamic conditions, the disk rapidly $(\lsim 1 $ yr) cools, the accretion rate drops, and the central core can grow only to $\approx 100 M_\odot$. Previous studies of the evolution of a cold ($T\simlt 100$ K), collapsing, metal-enriched core have convincingly demonstrated that the final result is a large number of sub--solar mass clumps that might instead eventually end up in a dense cluster  of low--  and intermediate--mass stars. 
To aggravate the situation, most of the disk becomes gravitationally unstable in $\approx 100$ yr, further quenching the accretion.
If instead the disk is born cold and turbulent, high accretion rates
can be maintained only as long as turbulence can be
supported. However, we find that turbulent energy is dissipated on a
time scale much shorter than the disk crossing time, thus almost
completely suppressing the initially large accretion rate onto the
core. These conclusions lead us to question the formation of the very
hot and dense core leading to direct collapse black hole seeds, as
found in the simulation of \cite{Mayer10}. 

Observationally, the existence of cold disks is supported by the detection of molecular emission in the center of (U)LIRG \citep{Sanders96, Scoville97, Downes98, Bryant99, Tacconi99, Downes07, Greve09}. The observations reveal molecular and dust disk-like structures with masses $10^{9-10}\;M_\odot$ within a few tens or hundreds of parsecs from the galaxy center (see e.g. Downes \& Eckart 2007). On other hand, there is no compelling evidence for (but also against) the presence of a hot component. We have to note however that the hot-disk phase is extremely short\footnote{In the original Mayer et al. (2010) simulation the hot phases lasts only for a few $\times 10^4$ yr, a duration decreased to $\sim 10^3$ yr if cooling is considered}, so that the probability to detect it is correspondingly low.

We suggest that the disagreement arises from the fact that either (a) the thermal structure of the disk is not properly described by the imposed polytropic equation of state (radiative cooling is not included in their refined simulations), or (b) turbulence dissipation is largely underestimated. We therefore recommend that future numerical work should aim at implementing a  proper treatment of the energy equation including cooling processes along with Adaptive Mesh Refinement methods to catch the physics of the inner pc at high spatial resolution.

\section*{Acknowledgments} 
We acknowledge useful discussions with M. Dotti, L. Mayer and M. Vietri.

\bibliographystyle{apj}
\bibliography{ref}

\newpage 
\label{lastpage} 
\end{document}